\documentclass[12pt]{article}
\usepackage{graphicx} 
\usepackage[english]{babel}
\usepackage{amsthm,amsmath,amsfonts,amssymb}
\usepackage[authoryear]{natbib}
\usepackage[colorlinks,citecolor=blue,urlcolor=blue]{hyperref}
\usepackage{ulem}
\usepackage{soul}
\soulregister\cite7
\soulregister\ref7
\soulregister\eqref7
\usepackage{pgfplots}
\pgfplotsset{compat=1.18}
\usepackage{booktabs}
\usepackage{xcolor}
\usepackage{xr}
\usepackage{adjustbox}
\usepackage{rotating}
\usepackage{subcaption}
\usepackage{authblk}
\usepackage{todonotes}
\theoremstyle{plain}

\newtheorem{theorem}{Theorem}[section]
\newtheorem{lemma}[theorem]{Lemma}
\newtheorem{proposition}[theorem]{Proposition}
\newtheorem{corollary}[theorem]{Corollary}
\newtheorem{assumption}{Assumption}
\newcounter{subassumption}[assumption]
\renewcommand{\thesubassumption}{\theassumption.\arabic{subassumption}}
\newenvironment{subassumptions}{
  \begin{list}{\thesubassumption}{%
    \usecounter{subassumption}%
    \setlength{\leftmargin}{1.5em}%
    \setlength{\labelsep}{0.5em}%
    \setlength{\itemsep}{0.5em}%
  }
}{
  \end{list}
}

\theoremstyle{remark}

\newtheorem{remark}[theorem]{Remark}

\newcommand{\rd}{\mathrm d}

\newcommand{\E}{\mathbb E}

\newcommand{\R}{\mathbb R}

\newcommand{\bs}{\mathbf}

\def \qmo{``}
\def \qmc{''}

\newcommand{\blind}{1} 

\date{\today}

\begin{document}

\def\spacingset#1{\renewcommand{\baselinestretch}%
{#1}\small\normalsize} \spacingset{1}

\if1\blind
{
  \title{Multivariate M-extremiles}
  \author{Luca Merlo \\
    Department of Human Sciences, Link Campus University\\
    and \\
    Valeria Bignozzi \\
    MEMOTEF Department, Sapienza University of Rome\\
    Lea Petrella \\
    MEMOTEF Department, Sapienza University of Rome\\
    and \\
    Nicola Salvati \\
    Department of Economics and Management, University of Pisa}
  \maketitle
} \fi

\if0\blind
{
  \bigskip
  \bigskip
  \bigskip
  \begin{center}
    {\LARGE\bf Robust multivariate extremiles}
\end{center}
  \medskip
} \fi


\begin{abstract}
\noindent
Extremiles represent an important class of distributional functionals defined through weighted squared loss functions, providing a valuable alternative to quantiles and expectiles. To date, their development has remained confined to the univariate setting. Here we introduce a robust multivariate extension of extremiles for random vectors by combining a directional weighting mechanism with Huber's multidimensional loss function. The proposed definition yields a family of directional extremiles indexed by a tuning parameter controlling the degree of robustness to outliers, with the quadratic-loss geometric extremile arising as a limiting case. We prove their main theoretical properties and introduce the notion of extremile sets. From an inferential perspective, we develop an efficient two-step procedure based on iteratively reweighted least squares, and derive asymptotic properties using the theory of Z-estimators. The finite-sample behavior of the proposed methodology is investigated through simulation experiments under Gaussian, contaminated, and heavy-tailed settings. An application to daily returns of three major cryptocurrencies illustrates its practical utility on financial data. 
\end{abstract}
\noindent \textbf{Keywords}: M-quantiles; Multivariate risk measures;  Generalized quantiles; Multivariate data; Robust estimation.



\newpage
\spacingset{1.8} 

\section{Introduction}

Characterizing the behavior of distributions in their tails is a central problem in statistics, with important implications in areas such as finance, insurance, environmental sciences, and risk management. Classical tools like quantiles provide interpretable summaries of distributional features and are widely used in practice. In quantitative risk management, for example, quantiles are directly linked to the Value-at-Risk measure, which is among the most widely used metrics for assessing financial risk.

Despite their widespread practical relevance, quantiles only exploit ordinal information, ignoring the magnitude of deviations from a given level. This limitation can be substantial in applications where the severity of extreme observations is as relevant as their occurrence. A prominent alternative addressing this issue is provided by \emph{expectiles}, introduced by \cite{newey1987asymmetric}, which arise as unique minimizers of asymmetric squared loss functions. Unlike quantiles, expectiles incorporate information on the magnitude of deviations through the quadratic loss and enjoy attractive theoretical properties, including elicitability and coherence \citep{artzner1999coherent,bellini2014generalized,delbaen2016risk}.

More recently, \cite{daouia2019extremiles} introduced a class of distributional functionals known as \emph{extremiles}, combining features of both quantiles and expectiles. Extremiles are defined through asymmetrically weighted squared loss functions that place greater emphasis on extreme observations while retaining a least-squares formulation. Compared with quantiles, extremiles account not only for the ordering of observations but also for the magnitude of tail realizations; in contrast to expectiles, their asymmetry is driven by the rank of an observation within the distribution rather than by the sign of its residual. This distinctive feature makes extremiles especially well suited for tail risk analysis, as they provide summaries that respond to the magnitude of extreme outcomes. Extremiles belong to the class of coherent distortion risk measures (see, e.g., \citealp{wang1996premium,acerbi2002spectral}) and thus provide a natural bridge between quantile-based and expectile-based approaches. Subsequent contributions have extended extremiles to regression settings \citep{daouia2022extremile,sun2026robust} and flexible modeling frameworks \citep{geng2024modelling}. The use of non-quadratic losses in the extremile literature is not entirely new. In particular, \cite{sun2026robust} employs Huber's loss in a univariate high-dimensional extremile regression model. More generally, \cite{debrauwer2024generalized} studies generalized univariate extremiles obtained by combining distortion weights with broad classes of loss functions. 

Despite these recent developments, extending extremiles to random vectors is inherently challenging because there is no natural total ordering in $\mathbb R^d$, $d\geq2$. A rich literature has proposed multivariate distributional summaries based on directional quantiles \citep{hallin2010multivariate,kong2012quantile}, depth concepts \citep{he1997convergence,serfling2002quantile}, and geometric constructions \citep{chaudhuri1996geometric,konen2022multivariate}. Analogous extensions have also been developed for multidimensional M-quantiles based on robust loss functions \citep{breckling1988m, kokic2002new, merlo2024unified}. Existing multivariate (M-)quantile-type summaries are not, however, designed to preserve the distinctive extremile mechanism, in which observations are weighted according to their position in the distribution. 
For multivariate expectiles, possible extensions rely on the notion of higher order elicitability as defined in \cite{fissler2016higher} or on geometric perspectives \citep{maume2017multivariate, herrmann2018multivariate}. While these methods retain information on the magnitude and direction of multivariate deviations, the quadratic loss function induces an unbounded score that is undesirable under contamination or heavy-tailed distributions. 

The present paper addresses this gap by introducing a new class of \emph{multivariate M-extremiles}. The proposed formulation extends the univariate definition of extremiles in \cite{daouia2019extremiles} in two distinct directions. The first is the directional nature that characterizes our extremile. A unit direction $\bs r$ is used to project the random vector $\bs Y$ onto $\bs r' \bs Y$, thereby inducing a univariate ordering of multivariate observations. The extremile distortion function then assigns different weights to each observation according to its position in this directional distribution. Second, the magnitude and geometry of multivariate deviations are determined through the multidimensional Huber’s function \citep{hampel2011robust} in place of the squared loss to provide robustness against heavy-tailed errors and outliers. Its tuning parameter $c \in [0,\infty)$ governs the transition between different multivariate extremile formulations and, for finite $c$, limits the contribution of large radial deviations through a bounded score. At $c=0$, the criterion reduces to the Euclidean norm and yields a spatial-type extremile, whereas, as $c\rightarrow\infty$, it approaches a quadratic criterion defining what we call the \emph{geometric extremile}. The latter provides the natural multivariate counterpart of the classical extremile of \cite{daouia2019extremiles}. More broadly, we also discuss the construction of multivariate M-extremiles associated with general radial convex loss functions. 
By varying the direction over the unit sphere, the proposed functional naturally generates extremile sets, thereby providing a geometric characterization of multivariate tail behavior. 

We establish the main theoretical properties of the proposed functional, including existence and uniqueness under mild conditions, symmetry relations, positive homogeneity and equivariance under data transformations. In addition, we show analytically that the resulting extremile sets are contained within the closed convex hull of the support of the underlying distribution, a desirable property for multivariate location functionals that ensures stability and interpretability. From an inferential perspective, we develop an efficient estimation procedure based on iteratively reweighted least squares and derive the asymptotic properties of our estimator using the theory of Z-estimators \citep{van2000asymptotic} and multivariate L-statistics. 

We illustrate the practical performance of the proposed method through simulation studies under Gaussian, contaminated, and heavy-tailed scenarios. From an empirical standpoint we demonstrate the usefulness of this method in the financial context through the analysis of daily returns of three major cryptocurrencies, namely Bitcoin, Ethereum and Binance Coin, from 2021 to 2026. These examples illustrate the ability of the proposed multivariate extremiles to capture tail features of multivariate distributions while maintaining robustness with respect to atypical observations.\\

The remainder of the paper is organized as follows. In Section \ref{sec:uni} we briefly review the univariate extremile. Section \ref{sec:meth} formally introduces the proposed multivariate M-extremile and the estimation procedure, and in Section \ref{sec:prop} we establish its theoretical properties. Section \ref{sec:sim} presents the simulation results, while Section \ref{sec:app} discusses the empirical application. Finally, Section \ref{sec:con} concludes. Additional simulation studies, computational metrics and all the proofs are provided in the Supplementary Materials. 

\section{Univariate extremiles}\label{sec:uni}
In this section, we briefly review the concept of univariate extremiles introduced in \cite{daouia2019extremiles}. Defined as weighted least squares analogues of quantiles, extremiles offer an alternative to both traditional quantiles and expectiles.

\begin{figure}[!h]
    \centering
\begin{tikzpicture}
\begin{axis}[
    domain=0.0001:0.9999,
    samples=200,
    xlabel={$t$},
    ylabel={$J_{\tau}(t)$},
    legend pos=north east,
    width=10cm,
    height=6cm,
    ymin=0, ymax=4,
    xmin=0, xmax=1
]
\addplot[blue, thick, solid] {(ln(0.5)/ln(0.9))*(1-x)^((ln(0.5)/ln(0.9))-1)};
\addlegendentry{$\tau=0.10$}
\addplot[red, thick,dashed] {(ln(0.5)/ln(0.75))*(1-x)^((ln(0.5)/ln(0.75))-1)};
\addlegendentry{$\tau=0.25$}

\addplot[green!60!black, thick, dotted] {(ln(0.5)/ln(0.5))*x^((ln(0.5)/ln(0.5))-1)};
\addlegendentry{$\tau=0.50$}

\addplot[gray, thick,dashdotted] {(ln(0.5)/ln(0.6))*x^((ln(0.5)/ln(0.6))-1)};
\addlegendentry{$\tau=0.60$}

\addplot[purple, thick, densely dashed] {(ln(0.5)/ln(0.9))*x^((ln(0.5)/ln(0.9))-1)};
\addlegendentry{$\tau=0.90$}

\end{axis}
\end{tikzpicture}
    \caption{The density function $J_\tau$ for $\tau = \{0.10, 0.25, 0.50, 0.60, 0.90\}$.}
    \label{fig:J_fun}
\end{figure}
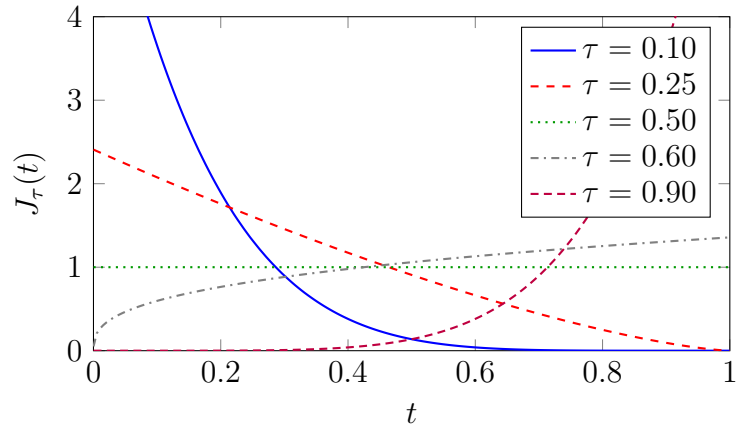

Let $Y$ denote a real-valued random variable with cumulative distribution function (CDF) $F_{Y}$ and finite mean. For any $\tau \in (0,1)$, the $\tau$-th extremile of $Y$, $ \xi_\tau$, is defined as
\begin{equation}\label{eq:extrem_u}
 \xi_\tau(Y) := \underset{ \xi \in \mathbb{R}}{\mathrm{argmin}} \, \E[J_\tau (F_{ Y}( Y)) (\lvert   Y - \xi  \rvert^2 - \lvert  Y  \rvert^2)],
\end{equation}
 where $J_\tau $ is the density function of the CDF $K_\tau $ given by 
\[
K_{\tau}(t) = 
\begin{cases} 
1 - (1 - t)^{r(1-\tau)} & 0 < \tau < \frac{1}{2}\\[6pt]
t^{r(\tau)} & \frac{1}{2} \leq \tau < 1,
\end{cases}
\]
with $r(\tau)  = \frac{\log\left(\frac{1}{2}\right)}{\log(\tau)}$ and $t\in[0,1]$. Hence,
\begin{equation}\label{eq:J}
J_{\tau}(t) = 
\begin{cases} 
  r(1-\tau)(1 - t)^{r(1-\tau)-1} & 0 < \tau < \frac{1}{2}\\[6pt]
r(\tau)t^{r(\tau)-1} & \frac{1}{2} \leq \tau < 1.
\end{cases}
\end{equation}
The  function $J_\tau$ (see Figure \ref{fig:J_fun}) assigns a weight to each observation according to its rank in the distribution. Therefore, the asymmetry is not induced by the sign of the residual $Y-\xi$, as happens for expectiles, but by the position of $Y$ within the distribution itself. In particular, when $\tau>1/2$, the weight function is strictly increasing and assigns larger weights to upper-tail observations; when $\tau<1/2$, it is strictly decreasing and emphasizes the lower tail, while for $\tau=1/2$, $J_{1/2}\equiv 1$. 

The extremile construction provides a direct link with quantiles
and expectiles. Replacing the squared loss in \eqref{eq:extrem_u} with the absolute value yields the  $\tau$-th quantile, while replacing $J_\tau$ with  $|\tau- \bs 1 (Y\leq \xi)|$ leads to the definition of expectiles.
 
Extremiles are by construction more sensitive to the magnitude of extreme values than quantiles and exploit information on both the distance to observations and their rank, while quantiles only depend on the frequency of observations below or above the predictor and not on their values. Unlike expectiles, extremiles benefit from various equivalent explicit representations. The definition in \eqref{eq:extrem_u} immediately highlights one of the main advantages of extremiles. Since the objective function is quadratic in $\xi$, the minimizer is unique whenever the expectation is finite, and, for continuous distributions,
\begin{equation}\label{eq:univ_extremile_closed}
\xi_\tau(Y) = \E[J_\tau (F_Y(Y)) Y] = \int_0^1 F^{-1}_Y (u)\,J_\tau(u)\,\rd u.
\end{equation}
The first equality shows that an extremile is a weighted mean with weights depending only on the rank of the observation. From the second equality in \eqref{eq:univ_extremile_closed} we observe that extremiles are $L$-functionals. Specifically, the $\tau$-th extremile is a weighted average of the quantile function $F_Y^{-1}$, with weights determined by $J_\tau$.
These weights shift continuously from the lower tail to the upper tail
as $\tau$ increases from $0$ to $1$.

Several basic properties follow from the above formulation. First, for $\tau=1/2$, $\xi_{1/2}(Y)=\E(Y)$. Second, extremiles also admit an interpretation in terms of expected maxima and minima. In particular, when $r(\tau)$ is an integer, upper extremiles correspond to expected maxima of independent copies of $Y$, with an analogous interpretation in terms of expected minima for lower extremiles. 
From a risk management perspective, for $\tau\geq 1/2$, extremiles belong to the class of coherent distortion risk measures (\citealp{artzner1999coherent,acerbi2002spectral}). A thorough description of basic properties of $\xi_\tau$ is given in \cite{daouia2019extremiles, daouia2022extremile}.

In the following section we introduce a new class of robust extremiles associated to multivariate variables.


\section{Multivariate robust extremiles}\label{sec:meth}

In this section, we extend the univariate extremile construction of
\cite{daouia2019extremiles} to random vectors in two distinct ways.
First,  we introduce a directional component by projecting the
random vector $\bs Y$ along a unit vector $\bs r$. The scalar projection
$\bs r'\bs Y$ provides a univariate ordering of multivariate observations
according to their exposure in direction $\bs r$, allowing the distortion
function $J_\tau$ to assign different weights to observations according to
this directional ordering. Second, we replace the quadratic loss underlying
classical extremiles with Huber's multidimensional M-function. While the
quadratic loss gives increasing influence to observations as their distance
from the extremile grows, the Huber loss becomes linear for sufficiently
large deviations, thereby limiting the impact of extreme observations and
providing robustness with respect to their magnitude.


Let $\bs Y$ denote a $d$-dimensional real-valued random vector,
$d\in\mathbb{N}$, and let $\bs r\in\mathcal{S}^{d-1}$ be a directional unit
vector, where
$
\mathcal{S}^{d-1}
=
\{\bs z\in\mathbb{R}^d:\|\bs z\|=1\}.
$
Under weak technical conditions (see Section \ref{sec:prop}), we define the
robust $\tau$-th multivariate extremile of $\bs Y$ in direction $\bs r$ as
\begin{equation}\label{eq:minextrem}
\boldsymbol\xi^c_{\tau,\bs r}(\bs Y)
=
\underset{\boldsymbol\xi\in\mathbb{R}^d}{\mathrm{argmin}}\,
\E\left[
J_\tau\left(F_{\bs r'\bs Y}(\bs r'\bs Y)\right)
\left(
\rho_c(\bs Y-\boldsymbol\xi)-\rho_c(\bs Y)
\right)
\right].
\end{equation}

To control sensitivity to the magnitude of multivariate deviations, we use
Huber's multidimensional M-function \citep{hampel2011robust},
\begin{equation}\label{eq:rhohuber}
\rho_c(\bs x)=
\begin{cases}
\dfrac{\|\bs x\|^2}{2c},
& \qquad \textrm{if } \|\bs x\|<c, \\[6pt]
\|\bs x\|-\dfrac{c}{2},
& \qquad \textrm{if } \|\bs x\|\geq c,
\end{cases}
\end{equation}
where $\bs x\in\mathbb{R}^d$, $c\in[0,\infty)$, and $\|\cdot\|$ denotes the
Euclidean norm. For $c=0$, the first branch of \eqref{eq:rhohuber} is never
attained, so that $\rho_0(\bs x)=\|\bs x\|$. The centering term
$\rho_c(\bs Y)$ does not affect the minimizer but reduces the moment
requirements needed to ensure its existence.

The three components entering \eqref{eq:minextrem} play distinct roles.
The level $\tau$ determines the degree and side of tail emphasis through
$J_\tau$, while the direction $\bs r$ determines the projection relative to
which observations are ordered and hence their relative distortion weights. For example, when restricted to the positive orthant, $\bs r$ also admits a useful
financial interpretation: its components are proportional to the weights of a fully invested long-only portfolio, so that the directional
ordering is induced by the corresponding portfolio exposure. This connection will be discussed further in Section \ref{sub:geomex}. Finally, the
tuning parameter $c$ acts through $\rho_c$ and controls robustness with
respect to the magnitude of multivariate deviations. Thus, $\tau$ and
$\bs r$ determine which observations receive greater emphasis, whereas $c$
controls their influence through the loss function.

Robust
multivariate extremiles can equivalently be characterized through the
first-order condition
\begin{equation}\label{eq:rextrem}
\Psi(\boldsymbol\xi)=\bs 0,
\qquad\textrm{where}\qquad
\Psi(\boldsymbol\xi)
=
\E\left[
J_\tau\left(F_{\bs r'\bs Y}(\bs r'\bs Y)\right)
\psi_c(\bs Y-\boldsymbol\xi)
\right],
\end{equation}
where
\begin{equation*}
\psi_c(\bs x)=
\begin{cases}
\dfrac{\bs x}{c},
& \qquad\textrm{if } \|\bs x\|<c, \\[6pt]
\dfrac{\bs x}{\|\bs x\|},
& \qquad\textrm{if } \|\bs x\|\geq c,
\end{cases}
\qquad \bs x\neq\bs 0,
\end{equation*}
and $\psi_c(\bs 0)=\bs 0$. 

The tuning parameter $c$ connects different multivariate extremile
constructions. At $c=0$, the criterion reduces to the Euclidean norm,
yielding a spatial-type extremile, and $\psi_0(\bs x)$ corresponds to the
spatial sign function. At the opposite extreme, as
$c\rightarrow\infty$, the family of robust multivariate extremiles converges,
under the conditions established in Section \ref{sec:prop}, to the quadratic
limiting extremile
\begin{equation}\label{eq:gextrem}
\boldsymbol\xi^\infty_{\tau,\bs r}(\bs Y)
=
\underset{\boldsymbol\xi\in\mathbb{R}^d}{\mathrm{argmin}}\,
\E\left[
J_\tau\left(F_{\bs r'\bs Y}(\bs r'\bs Y)\right)
\left(
\|\bs Y-\boldsymbol\xi\|^2-\|\bs Y\|^2
\right)
\right],
\end{equation}
which we refer to as the \textit{geometric extremile} of $\bs Y$ in direction
$\bs r$ at level $\tau$. The terminology recalls geometric quantiles and geometric expectiles
\citep{chaudhuri1996geometric,herrmann2018multivariate}, although the role of
the direction is different: here $\bs r$ enters only through the directional
distortion weight and not through the loss function.

For $d=1$ and $r=1$, \eqref{eq:minextrem} reduces to the univariate robust
extremile of \cite{sun2026robust}, while the geometric extremile
\eqref{eq:gextrem} recovers the univariate extremile of
\cite{daouia2019extremiles}. Our construction is also related to the
multivariate M-quantile framework of \cite{kokic2002new}, which similarly
employs a Huber-type criterion. In our setting, however, the directional
component enters exclusively through the distortion weight
$J_\tau(F_{\bs r'\bs Y}(\bs r'\bs Y))$, rather than through an asymmetric
residual-dependent loss.

Finally, varying the direction over the unit sphere defines, for fixed
$\tau$ and $c$, the directional M-extremile set
\begin{equation*}\label{eq:mqcon}
\mathcal{A}^c_\tau(\bs Y)
:=
\left\{
\boldsymbol\xi^c_{\tau,\bs r}(\bs Y)
\,\big|\,
\bs r\in\mathcal{S}^{d-1}
\right\}.
\end{equation*}
Properties of $\boldsymbol\xi^c_{\tau,\bs r} (\bs Y)$ and of the corresponding sets are studied in the following section and in the numerical analyses.

\section{Properties}\label{sec:prop}
We provide some desirable properties satisfied by the multivariate M-extremiles {for $d \geq 2$}. 
We begin by establishing the mathematical well-posedness of $\boldsymbol\xi^c_{\tau,\bs r} (\bs Y)$. As introduced in Section~\ref{sec:meth}, we distinguish between the robust multivariate extremiles $\boldsymbol\xi^c_{\tau,\bs r}(\bs Y)$, defined through the Huber loss for finite $c$, and the geometric extremile $\boldsymbol\xi^\infty_{\tau,\bs r}(\bs Y)$, defined through the quadratic criterion in \eqref{eq:gextrem}. We first establish existence and uniqueness for both objects. We then show that the geometric extremile is indeed obtained as the limit of the robust family as $c\to\infty$.

The following conditions, which closely follow the framework of \cite{chaudhuri1996geometric}, are sufficient for our purposes and will be assumed throughout the paper.

\begin{assumption}\label{ass:mod1} Throughout the paper we assume the following conditions hold:
\begin{subassumptions}
\item\label{ass:samp_mod_reg_a} The distribution of $\bs Y$ is not completely supported on a straight line in $\mathbb{R}^d$.

\item\label{ass:samp_mod_reg_b} $\E[ \| \bs Y  \|] < \infty$.

\item\label{ass:samp_mod_reg_c} The distribution of $\bs r' \bs Y$ is continuous.
\end{subassumptions}
\end{assumption}

Assumption \ref{ass:samp_mod_reg_a} is standard to ensure uniqueness for multivariate-type quantiles, such as geometric or spatial quantiles, due to the fact that $ \| \cdot \|$ is convex but not strictly convex in general. Assumption \ref{ass:samp_mod_reg_b} postulates that the distribution of $\bs Y$ has finite first moment; because of the recentering term in \eqref{eq:minextrem}, this is a sufficient condition needed for the geometric extremile only to make the objective function coercive and ensure that the minimizer exists. Assumption \ref{ass:samp_mod_reg_c} prevents the projection on $\bs r$ from being degenerate and the weight $J_\tau$ to be identically zero. Together, these conditions guarantee both existence and uniqueness of $\boldsymbol \xi^c_{\tau,\bs r} (\bs Y)$, as shown in the following proposition.

\begin{proposition}\label{prop:ex}
\underline{Existence and Uniqueness}: For any $\tau\in(0,1)$ and $\bs r\in\mathcal S^{d-1}$, the multivariate extremile $\boldsymbol\xi^c_{\tau,\bs r}(\bs Y)$ exists and is unique for every finite $c\geq 0$. Moreover, the geometric extremile $\boldsymbol\xi^\infty_{\tau,\bs r}(\bs Y)$ defined in \eqref{eq:gextrem} exists and is unique.
\end{proposition}

\begin{remark} \underline{Radial convex losses}: Proposition \ref{prop:ex} is stated for the multidimensional Huber loss. 
More generally, multivariate M-extremiles can be defined by replacing $\rho_c$ in \eqref{eq:rhohuber} with a radial loss of the form $\rho_\phi(\bs x)=\phi(\|\bs x\|)$, for suitable convex and increasing functions $\phi$. Examples include power loss functions $\phi(t)=t^p$, $p\geq 1$, or exponential functions $\phi(t) = \exp(at) -1$, $a>0$. Under appropriate conditions on $\phi$ and the distribution of $\bs Y$ \citep{konen2022multivariate}, existence and uniqueness continue to hold. The corresponding statement and proof are provided in Section S4 of the Supplementary Materials.
\end{remark}

The next result shows that the geometric extremile coincides with $\boldsymbol\xi^\infty_{\tau,\bs r}(\bs Y)$ by taking the limit of \eqref{eq:minextrem} as $c\to\infty$.

{\begin{proposition}\label{prop:limiting}
\underline{Convergence to the geometric extremile}.
For any $\tau\in(0,1)$ and $\bs r\in\mathcal S^{d-1}$,
\[
\lim_{c\to \infty}\boldsymbol\xi^c_{\tau,\bs r}(\bs Y)
=
\boldsymbol\xi^\infty_{\tau,\bs r}(\bs Y).
\]
\end{proposition}}

Having established existence and uniqueness, we now investigate some structural properties of multivariate extremiles. In what follows, whenever a property holds for both finite $c$ and the geometric extremile case, we use the notation $c\in[0,\infty]$, with $c=\infty$ referring to the geometric extremile $\boldsymbol\xi^\infty_{\tau,\bs r}(\bs Y)$. For notational consistency, we denote $\rho_{\infty}(\bs x)=\frac{1}{2}||\bs x||^2$ and $\psi_{\infty}(\bs x)=\bs x$, for any $\bs x\in\R^d$. Furthermore, for $c=\infty$,  any $\tau\in(0,1)$ and $\bs r\in\mathcal{S}^{d-1}$, minimizing \eqref{eq:gextrem} yields the following  closed-form representation
\begin{equation}
\label{eq:gextrem_closed}
\boldsymbol\xi^\infty_{\tau,\bs r}(\bs Y)
=
\E\left[
J_\tau\left(
F_{\bs r'\bs Y}(\bs r'\bs Y)
\right)\bs Y
\right].
\end{equation}
We begin with the following symmetry property.

\begin{proposition}\label{prop:symmetry} \underline{Symmetry}: For any $\tau\in(0,1),~ c\in[0,\infty]$ and direction $\bs r\in \mathcal{S}^{d-1}$, 
\begin{equation}\label{eq:symm}\boldsymbol \xi^c_{1-\tau, \bs r}(\bs Y) = \boldsymbol \xi^c_{\tau, -\bs r}(\bs Y).
\end{equation}
\end{proposition}

This result reflects the inherent symmetry of the weighting function $J_\tau$. In practice, this implies that if one is interested in $\boldsymbol \xi_{1-\tau, \bs r}(\bs Y)$, with $\tau \in (0,1/2]$, one simply sets $\boldsymbol \xi_{1-\tau, \bs r}(\bs Y) = \boldsymbol \xi_{\tau, -\bs r}(\bs Y)$. 

We turn our attention to translation equivariance and orthogonal equivariance. Positive homogeneity is  established for the cases $c=0$ and $c=\infty$. To do so, we first show that the density function $J_\tau$ in \eqref{eq:J} is invariant to location-scale and orthogonal transformations of the vector $\bs Y$.
\begin{lemma}\label{Lem:trans} Let $\bs r \in \mathcal{S}^{d-1}$, $\bs W=s\bs Y+\bs b$, where $s>0,$ and $\bs b\in\mathbb{R}^d$. Let $\bs A$ be a $d\times d$-orthogonal matrix, $\bs W^+ = \bs A \bs Y$ and $\bs v=\bs A\bs r$. Then,
 \begin{align}
   &(i)  \qquad J_{\tau}(F_{\bs r' \bs Y }(\bs r'\bs Y))=J_\tau(F_{\bs r'\bs W}(\bs r'\bs W)),\label{eq:transf}\\
   &(ii) \qquad J_\tau(F_{\bs r'\bs Y}(\bs r'\bs Y))=J_{\tau}(F_{\bs v'\bs W^+}(\bs v'\bs W^+)).
\end{align}
\end{lemma}
\begin{proof}
Let $\tilde Y=\bs r'\bs Y$, $\tilde W=\bs r'\bs W$ and $\tilde W^+=\bs v'\bs W^+$. To prove (i), we first note that the projection of $\bs W$ in the direction $\bs r$ is given by
$$
\tilde W=\bs r'\bs W=\bs r's\bs Y+\bs r'\bs b=s\tilde Y+\bs r'\bs b.
$$
For any $t\in\mathbb{R}$, the CDF of $\tilde W$ is given by
\begin{equation*}
F_{\tilde W}(t)=\mathbb{P}(s\tilde Y\leq t-\bs r'\bs b)=\mathbb{P}\left(\tilde Y\leq \frac{t}{s}-\bs r'\frac{\bs b}{s}\right)=F_{\tilde Y}\left(\frac{t}{s}-\bs r'\frac{\bs b}{s}\right)
\end{equation*}
from which $F_{\tilde Y}(\tilde Y)=F_{\tilde W}(\tilde W)$, and Equation \eqref{eq:transf} follows.

To prove (ii), note that $\tilde W^+=\bs v'\bs W^+=(\bs A\bs r)'\bs A\bs Y=\bs r'\bs A'\bs A\bs  Y=\bs r'\bs  Y=\tilde Y$ with $ \| \bs v \|= \| \bs r \|$. Since $\tilde W^+ = \tilde Y$, their distribution functions are identical.
\end{proof}

Given Lemma \ref{Lem:trans}, we derive the following fundamental properties that ensure extremiles are well-behaved under common data transformations.

\begin{proposition}\label{prop:equivariance}For any $\tau\in(0,1)$ and direction $\bs r\in \mathcal{S}^{d-1}$
\begin{enumerate}
\item[I)] \underline{Location equivariance}: For any $\bs b\in\R^d$ and $c\in[0,\infty]$, $$\boldsymbol \xi^c_{\tau,\bs r}(\bs Y+ \bs b)=\xi^c_{\tau,\bs r}(\bs Y)+\bs b.$$
\item[II)] \underline{Orthogonal transformation equivariance}: Let $A$ be a $d\times d$-orthogonal matrix and $\bs v=\bs A\bs r$, then for any $c\in[0,\infty]$
\begin{equation}\label{eq:orth}
\boldsymbol \xi^c_{\tau,\bs v}(\bs A\bs Y)=\bs A\boldsymbol \xi^c_{\tau,\bs r}(\bs Y).
\end{equation}
\item[III)] \underline{Scaling equivariance}: Let $s > 0$ with $c \in [0,\infty)$. Then
\begin{equation*}\label{eq:scale}
\boldsymbol \xi^c_{\tau,\bs r}(s\bs Y)=s\boldsymbol \xi^{c/s}_{\tau,\bs r}(\bs Y).  
\end{equation*}
\item[IV)]\underline{Positive homogeneity}: For any $s > 0$ and  $c=0$ or $c=\infty$, one has
\begin{equation}
 \boldsymbol \xi^c_{\tau,\bs r}(s\bs Y)=s\boldsymbol \xi^{c}_{\tau,\bs r}(\bs Y).  
\end{equation}
\end{enumerate}
\end{proposition}

Location equivariance guarantees that shifts in the data are reflected one-to-one in the functional, while orthogonal equivariance ensures that rotating the coordinate system simply rotates the extremile. Scaling equivariance shows that scaling the data requires a corresponding scaling of the tuning constant. Indeed, for $0 < c < \infty$, the Huber's M-function $\rho_c$ in \eqref{eq:rhohuber} is not positively homogeneous. However, positive homogeneity holds for the  cases $c=0$ and $c=\infty$.


From rotational equivariance, Corollary \ref{Cor:sign} and Proposition \ref{prop:index} below explore the symmetry properties of $\boldsymbol \xi^c_{\tau, \bs r}(\bs Y)$.

\begin{corollary}\label{Cor:sign} \underline{Vector sign symmetry}: For any $\tau\in(0,1)$, direction $\bs r\in \mathcal{S}^{d-1}$ and $c\in[0,\infty]$ $$\boldsymbol \xi^c_{\tau, \bs r}(-\bs Y) = -\boldsymbol \xi^c_{\tau, -\bs r}(\bs Y).$$
\end{corollary}

Corollary \ref{Cor:sign} describes how the functional behaves under reflection of the entire random vector. In particular, changing the sign of the data corresponds to changing the sign of the extremile and reversing the direction. Combining Proposition \ref{prop:symmetry} and Corollary \ref{Cor:sign}, it follows immediately that $\boldsymbol \xi^c_{1-\tau, \bs r}(\bs Y) = -\boldsymbol \xi^c_{\tau, \bs r}(-\bs Y)$.

For a symmetric distribution with center $\bs a \in \mathbb{R}^d$ in the sense that the random vectors $\bs Y - \bs a$ and $\bs a - \bs Y$ are identically distributed, the extremile obeys the following symmetry relation.

\begin{proposition}\label{prop:index} \underline{Index sign symmetry}: Suppose that the distribution of $\bs Y$ is symmetric with respect to the reflection about $\boldsymbol a\in \mathbb{R}^d$, then {for any $\tau\in(0,1)$, direction $\bs r\in \mathcal{S}^{d-1}$ and $c\in[0,\infty]$}$$\boldsymbol \xi^c_{\tau, -\bs r}(\bs Y) = 2\bs a -\boldsymbol \xi^c_{\tau, \bs r}(\bs Y).$$
\end{proposition}

From Proposition \ref{prop:index} follows the closed-form expression for the extremile at $\tau = 1/2$, $\boldsymbol \xi^c_{1/2, \bs r} = \bs a$ for any direction $\bs r \in \mathcal{S}^{d-1}$. More in general, $\bs a = \frac{1}{2}(\boldsymbol \xi^c_{\tau, \bs r} + \boldsymbol \xi^c_{\tau, -\bs r})$ for all $\bs r \in \mathcal{S}^{d-1}$.\\


We consider a fundamental requirement for a multivariate location functional, namely that it lies within the closed convex hull of the support of $\mathbf{Y}$. This property is referred to as internality.

\begin{proposition}\label{prop:hull} \underline{Convex hull internality}: For any $\tau \in (0,1)$, direction $\bs r \in \mathcal{S}^{d-1}$ and $c\in[0,\infty]$, the extremile  ${\boldsymbol \xi}^c_{\tau, \bs r}(\bs Y)$ lies inside the closed convex hull of the distribution of $\bs Y$.
\end{proposition}

At extreme values of $\tau$, this property is satisfied neither by the multivariate M-quantiles of \cite{breckling1988m} nor by geometric-type quantiles and expectiles \citep{girard2017intriguing}. While the multivariate M-quantiles proposed by \cite{kokic2002new} appear to satisfy this empirically, a formal proof remains absent. 
See the Supplementary Materials for a comparative analysis with the proposed multivariate extremiles. 

\subsection{Geometric extremiles}\label{sub:geomex}
We restrict the attention to the geometric case, $c=\infty$, for which the quadratic structure yields additional properties with a direct financial interpretation. First, note that geometric extremiles  are consistent with the univariate extremile of the corresponding directional projection. Indeed, for any $\tau\in(0,1)$ and $\bs r\in\mathcal{S}^{d-1}$, 
\begin{equation}
\label{eq:gextrem_proj}
\qquad \bs r'\boldsymbol\xi^\infty_{\tau,\bs r}(\bs Y)
=
\xi_\tau(\bs r'\bs Y).
\end{equation}
Equation \eqref{eq:gextrem_proj} provides a direct link with the portfolio interpretation introduced in Section \ref{sec:meth}. When $\bs r \in \mathcal{S}_+^{d-1}$, where $\mathcal{S}_+^{d-1} = \{ \bs z \in \mathbb{R}^d_+: \| z \| = 1\}$ denotes the portion of the $(d-1)$-dimensional unit sphere in the positive orthant, the elements of the normalized direction vector $\bs w = \frac{\bs r}{\bs1'\bs r}$ correspond to the weights of a long-only, fully invested portfolio. Furthermore, writing $\bs r = (\bs1'\bs r)\bs w$, positive homogeneity of the univariate extremile gives
\begin{equation*}
\bs w'
\boldsymbol\xi^\infty_{\tau,\bs w}(\bs Y)
=
\xi_\tau(\bs w' \bs Y).
\end{equation*}

Thus, aggregating the components of the geometric extremile using the
portfolio weights exactly recovers the univariate extremile of the
corresponding portfolio loss, which we label as \emph{directional extremile}.

Finally, we study the properties of the extremile sets $\mathcal{A}^\infty_\tau (\bs Y)$, including convexity and subadditivity of the corresponding regions. 

\begin{proposition}\label{prop:convex} \underline{Convexity of geometric extremile sets}: Assume that $\bs Y$ is such that $F_{\bs r' \bs Y}$ is continuous for every $\bs r \in \mathcal{S}^{d-1}$. {For $\tau \in (0,1)$}, define the associated geometric extremile region by 
 \[
\mathcal R^\infty_\tau(\bs Y)
=
\begin{cases}
\displaystyle
\bigcap_{\bs r\in\mathcal S^{d-1}}
\left\{
\bs y\in\mathbb R^d:
\bs r^\prime \bs y
\geq
\bs r^\prime \boldsymbol\xi^\infty_{\tau,\bs r}(\bs Y)
\right\},
& \tau\in(0,1/2],\\[12pt]
\displaystyle
\bigcap_{\bs r\in\mathcal S^{d-1}}
\left\{
\bs y\in\mathbb R^d:
\bs r^\prime \bs y
\leq
\bs r^\prime \boldsymbol\xi^\infty_{\tau,\bs r}(\bs Y)
\right\},
& \tau\in(1/2,1)
\end{cases}
\]
 and its boundary with $\partial \mathcal{R}^\infty_\tau (\bs Y)$.
 Then, $\mathcal{R}^\infty_\tau (\bs Y)$ coincides with the convex hull of $\mathcal{A}^\infty_\tau (\bs Y)$ and $\partial \mathcal{R}^\infty_\tau (\bs Y)$ with $\mathcal{A}^\infty_\tau (\bs Y)$.
 \end{proposition}

With respect to the existing literature, the regions $\mathcal{R}^\infty_\tau (\bs Y)$ coincides with the CECH$^*$ regions studied in \cite{dyckerhoff2011weighted} at level $\alpha = \frac{\log(\max\{\tau,1-\tau\})}{\log1/2}$. We conclude this section by investigating the multivariate sub-additivity of extremiles. Although univariate extremiles are inherently sub-additive, the lack of a canonical ordering in $\mathbb{R}^d$, $d \geq 2$, complicates the multivariate extension of this property. Here we address this by characterizing sub-additivity based on set-based inclusions along the same lines of \cite{herrmann2018multivariate}. 

\begin{corollary}\label{Cor:sub} \underline{Subadditivity of geometric extremile regions}: Let $\bs X$ and $\bs Y$ two $d$-dimensional random vectors such that $\bs X$, $\bs Y$ and $\bs X + \bs Y$ satisfy Assumption \ref{ass:samp_mod_reg_c} for every $\bs r \in \mathcal{S}^{d-1}$. For $\tau \in (0, 1)$, define the set
\begin{equation*}
\mathcal{A}^\infty_\tau (\bs X, \bs Y) = \{\boldsymbol \xi^\infty_{\tau, \bs r} (\bs X) + \boldsymbol \xi^\infty_{\tau, \bs r} (\bs Y) \, \big| \, \bs r \in \mathcal{S}^{d-1} \}
\end{equation*}
and $\mathcal{R}^\infty_\tau (\bs X, \bs Y)$ the corresponding extremile region
\[
\mathcal{R}^\infty_\tau(\bs X,\bs Y)
=
\begin{cases}
\displaystyle
\bigcap_{\bs r\in\mathcal S^{d-1}}
\left\{
\bs z\in\mathbb R^d:
\bs r'\bs z
\geq
\bs r'
\left(
\boldsymbol\xi^\infty_{\tau,\bs r}(\bs X)
+
\boldsymbol\xi^\infty_{\tau,\bs r}(\bs Y)
\right)
\right\},
& \tau\in(0,1/2],\\[12pt]
\displaystyle
\bigcap_{\bs r\in\mathcal S^{d-1}}
\left\{
\bs z\in\mathbb R^d:
\bs r'\bs z
\leq
\bs r'
\left(
\boldsymbol\xi^\infty_{\tau,\bs r}(\bs X)
+
\boldsymbol\xi^\infty_{\tau,\bs r}(\bs Y)
\right)
\right\},
& \tau\in(1/2,1).
\end{cases}
\]
Then, the multivariate geometric extremile regions satify the following subadditivity property:
\begin{equation*}
\mathcal{R}^\infty_\tau (\bs X + \bs Y) \subseteq \mathcal{R}^\infty_\tau (\bs X, \bs Y),
\end{equation*}
for all $\tau \in (0, 1)$.
\end{corollary}

We stress that the above subadditivity result is established only for the geometric case $c=\infty$. For finite values of $c$, the lack of positive homogeneity of Huber's loss prevents a direct extension of the same argument, and set-subadditivity does not hold in general. Investigating precise conditions under which analogous inclusion results may be recovered for finite $c$ is left for future research.

\subsection{Estimation}\label{sub:est}
In this section, we discuss the estimation of $\boldsymbol \xi^c_{\tau, \bs r}$, provide a closed-form solution to the geometric extremile minimization problem in \eqref{eq:gextrem} and then establish asymptotic properties of our estimator.

As in \cite{daouia2019extremiles, daouia2022extremile}, we suggest a two-step procedure. For a given $\tau$, direction $\bs r$ and $c$, we first estimate the empirical CDF of the projection $\tilde Y$. An estimate of $\boldsymbol \xi^c_{\tau, \bs r}$ is then obtained by solving \eqref{eq:minextrem} via iteratively reweighted least squares (IRLS) where we substitute $F_{\tilde Y}$ with the estimated CDF $\hat F_{\tilde Y}$. Specifically, let $\bs Y_i$, $i = 1,\dots,n$, denote a random sample of size $n$, the estimator of the extremile in \eqref{eq:minextrem}, $\hat{\boldsymbol \xi}^c_{\tau, \bs r}$, is defined as the solution of the empirical f.o.c. $\Psi_n(\boldsymbol \xi) = \bs 0$ with,
\begin{equation}\label{eq:emp_foc}
\Psi_n(\boldsymbol \xi) = \frac{1}{n} \sum_{i=1}^n J_\tau(\hat F_{\tilde Y}(\tilde Y_i)) \psi_c(\bs Y_i - \boldsymbol \xi).  
\end{equation}
The IRLS is executed iteratively updating the weighting scheme induced by $\psi_c(\bs Y_i - \boldsymbol \xi)$ until convergence is achieved. Convergence of the algorithm is reached when the norm of the difference between the extremiles at two consecutive iterations falls below a threshold $\epsilon>0$ and it is initialized using the empirical weighted mean with weights $J_\tau(\hat F_{\tilde Y}(\tilde Y_i))$. While we currently use the empirical distribution function to estimate the CDF, other methods like rank-based or kernel-density estimation are also viable options. This two-step strategy mirrors the univariate extremile estimation strategy while remaining computationally tractable in the multivariate setting.

Typically, the extremile estimator solution to \eqref{eq:emp_foc} is not available in closed form. For generic distributions and finite $c$, an explicit formula exists only in limiting cases. At $\tau = 1/2$, $J_\tau \equiv 1$, if the distribution of $\bs Y$ is centrally symmetric about $\bs a$, then (by Proposition \ref{prop:index}) $\hat{\boldsymbol \xi}^c_{1/2, \bs r}$ is an estimator for $\bs a$. For $c = \infty$, the estimating equation $\Psi_n(\boldsymbol \xi)$ becomes linear in $\boldsymbol \xi$, hence the solution is an empirical weighted average of \eqref{eq:gextrem_closed} similar to the univariate setting, $\hat{\boldsymbol \xi}^{\infty}_{\tau, \bs r} =\frac{\sum_{i=1}^n J_\tau(\hat F_{\tilde Y}(\tilde Y_i)) \bs Y_i}{\sum_{i=1}^n J_\tau(\hat F_{\tilde Y}(\tilde Y_i))}$. If in addition $\bs Y$ is elliptical, the extremile is in closed-form up to a scalar that can be computed from a univariate integral, as illustrated in the following proposition.

\begin{proposition}\label{prop:closed}
Let $c = \infty$, if $\bs Y$ is elliptical with center $\boldsymbol \mu$, positive-definite scatter matrix $\bs \Sigma$ and generator function $\varphi$, then
\begin{equation}\label{eq:closed}
\boldsymbol \xi^\infty_{\tau, \bs r} = \boldsymbol \mu + \kappa_\tau \frac{\bs \Sigma \bs r}{\sqrt{\bs r' \bs \Sigma \bs r}},    
\end{equation}
where $\kappa_\tau$ is a scalar value that depends only on the one-dimensional standardized projection of $\bs Y$, that is
\begin{equation*}
\kappa_\tau = \E[J_\tau(F_Z (Z)) Z] = \int_0^1 F^{-1}_Z(u) J_\tau (u) \rd u,
\end{equation*}
with $Z = \bs r'(\bs Y - \boldsymbol \mu)/\sqrt{\bs r' \bs \Sigma \bs r}$. 
\end{proposition}

By Eq. \eqref{eq:closed}, the central extremile is just the expectation of $\bs Y$. For $\tau \neq 1/2$, the extremile in direction $\bs r$ is obtained by taking the center $\boldsymbol \mu$, then moving along the projection direction $\bs r$ warped by the scatter matrix $\bs \Sigma$ 
determined according to the shape of the elliptical distribution, with signed distance given by $\kappa_\tau$. This scalar controls whether one moves outward in the positive or negative tail direction, and how far from $\boldsymbol \mu$ the extremile lies. By varying $\bs r$ over $\mathcal{S}^{d-1}$, this means the extremile region has an elliptically shaped contour centered at $\boldsymbol \mu$. An estimate of the extremile can be obtained directly from \eqref{eq:closed} by plugging-in an estimate of $\boldsymbol \mu$ and $\bs \Sigma$, and computing the integral for $\kappa_\tau$ numerically.

\subsection{Asymptotic properties}\label{sub:asy}
This section provides consistency and asymptotic normality of the proposed estimator in \eqref{eq:emp_foc} for $\boldsymbol \xi^c_{\tau, \bs r}$. 

To discuss consistency of $\hat{\boldsymbol \xi}^c_{\tau, \bs r}(\bs Y)$, we begin from the empirical f.o.c. $\Psi_n (\boldsymbol \xi)$ and proceed by using the theory of Z-estimation \citep{van2000asymptotic}.

\begin{proposition}\label{prop:consistency} \underline{Consistency}: For any $\tau \in (0,1)$, direction $\bs r \in \mathcal{S}^{d-1}$ and $c \in [0,\infty]$, the estimator $\hat{\boldsymbol \xi}^c_{\tau, \bs r}(\bs Y)$ converges in probability to $\boldsymbol \xi^c_{\tau, \bs r}(\bs Y)$.    
\end{proposition}

We finally prove asymptotic normality of $\hat{\boldsymbol \xi}^c_{\tau, \bs r}(\bs Y)$ for fixed values of $c$, $\tau$ and $\bs r$. Before stating the theorem, we need the following additional assumptions.
\begin{assumption}\label{ass:mod2} Throughout the rest of this section we assume the following conditions hold:
\begin{subassumptions}
\item\label{ass:Leb} The distribution of the random vector $\bs Y$ is absolutely continuous with respect to the Lebesgue measure on $\mathbb{R}^d$.

\item\label{ass:nabla} Consider the following matrix-valued function: 
\begin{equation*}
h_c (\bs x) = \begin{cases} 
        \frac{1}{c} \bs I & \qquad\textrm{if }  \| \bs x  \| < c \\
          \frac{1}{\Vert \bs{x} \Vert} \big( \bs I - \frac{\bs{x} \bs{x}'}{ \| \bs x  \|^2} \big) &\qquad\textrm{if } \| \bs x  \| \geq c. \end{cases}
\end{equation*}

\item\label{ass:invA} Assume that $\Psi(\boldsymbol \xi)$ is differentiable at $\boldsymbol \xi_{\tau, \bs r}$ with Jacobian matrix
\begin{equation}\label{eq:A}
\bs A_c (\boldsymbol\xi) = -\nabla_{\boldsymbol\xi} \Psi(\boldsymbol \xi) = -\nabla_{\boldsymbol\xi} \E[W  \psi_c (\bs Y - \boldsymbol \xi)], \qquad W = J_\tau(F_{\tilde Y}(\tilde Y))
\end{equation}
and $\bs A_c$ is non-singular.
\end{subassumptions}
\end{assumption}

Assumptions \ref{ass:Leb}-\ref{ass:invA} are standard in robust estimation theory \citep{hallin2010multivariate, hampel2011robust}, and ensure the existence and invertibility of $\bs A_c$.

To establish asymptotic normality, we first derive a linear Bahadur-type representation \citep{bahadur1966note} for the estimator $\hat{\boldsymbol \xi}^c_{\tau, \bs r}$ which allows us to characterize its limiting distribution. The analysis is partitioned into two scenarios: the case $c>0$ is detailed in Theorem \ref{prop:asynorm}, while the case  $c=0$ is addressed in Corollary \ref{cor:asynorm}.

\begin{theorem}\label{prop:asynorm} \underline{Asymptotic normality}: Set $0<c<\infty$ and let $\boldsymbol \xi^c_{\tau, \bs r}$ be the unique solution to $\Psi(\boldsymbol \xi) = \bs 0$. Then,
\begin{equation*}
\sqrt{n}(\hat{\boldsymbol \xi}^c_{\tau, \bs r} - \boldsymbol\xi^c_{\tau, \bs r}) \overset{d}{\rightarrow} \mathcal{N}(\bs 0, \bs A_c^{-1} \bs B_c \bs A_c^{-1}),   
\end{equation*}
where $\bs A_c$ is the Jacobian in \eqref{eq:A} evaluated at $\boldsymbol\xi^c_{\tau, \bs r}$ and $\bs B_c$ is given by
\begin{equation*}
\bs B_c = \int_0^1 J^2_\tau (w) \bs \Sigma_c(w)  \rd w + \int_0^1 \int_0^1 (s \wedge t - st) J_\tau(s) J_\tau(t) \rd \boldsymbol \alpha_c(s) \rd \boldsymbol \alpha_c(t)',
\end{equation*}
with $\bs Z_c = \psi_c(\bs Y - \boldsymbol\xi^c_{\tau, \bs r})$, $\boldsymbol \alpha_c(w) = \E[ \bs Z_c \mid W = w ]$ admits a component-wise right continuous function of bounded variation on $[0,1]$, $W = F_{\tilde Y} (\tilde Y) \sim U(0,1)$, and $\bs \Sigma_c(w) = \E[(\bs Z_c - \boldsymbol \alpha_c(w))(\bs Z_c - \boldsymbol \alpha_c(w))' \mid W = w]$.
\end{theorem}

The same asymptotic normality result extends to $c=\infty$ assuming also $\E[\| \bs Y \|^2] < \infty$, with $\psi_\infty (\bs x) = \bs x$, $\mathbf A_\infty=\mathbf I_d$ and $\mathbf B_\infty$ obtained from the expression in Theorem \ref{prop:asynorm} by setting $\mathbf Z_\infty= \mathbf Y- \boldsymbol\xi_{\tau,\mathbf r}^{\,\infty}$.


We now focus on the case $c=0$. Before stating the corollary, it is useful to observe that $\bs A_0$ in \eqref{eq:A} exists finitely only if $\bs Y$ admits bounded density in a neighborhood of $\boldsymbol \xi^0_{\tau,\bs r}$ \citep{niemiro1992asymptotics}. 
This condition is also needed to ensure that the empirical Jacobian $\bs A_n$ converges to the true one.

\begin{corollary}\label{cor:asynorm}
Assume that the density of $\bs Y$ is bounded in a neighborhood of $\boldsymbol \xi^0_{\tau,\bs r}$. Define $\Delta_n = \hat{\boldsymbol \xi}^0_{\tau, \bs r} - \boldsymbol \xi^0_{\tau,\bs r}$ and $\bs A_n (\boldsymbol \xi^\star) = \frac{1}{n} \sum_{i=1}^n J_\tau(\hat F_{\tilde Y}(\tilde Y_i) h_0(\bs Y_i - \boldsymbol \xi^0_{\tau,\bs r})$. Then the convergence rate of the remainder $R_n(\Delta_n)$ of the following linearization of $\Psi_n$ around $\boldsymbol \xi^0_{\tau,\bs r}$
\begin{equation*}
\Psi_n(\hat{\boldsymbol \xi}^0_{\tau, \bs r}) = \Psi_n (\boldsymbol \xi^0_{\tau,\bs r}) - \bs A_n (\boldsymbol \xi^0_{\tau,\bs r}) \Delta_n + R_n (\Delta_n)    
\end{equation*}
is 
\begin{equation*}
R_n(\Delta_n) = \begin{cases} 
        O_p(n^{-1}) & \qquad\textrm{if } d>2 \\
        O_p(\log(n)/n) &\qquad\textrm{if } d = 2. \end{cases}
\end{equation*}
So in all cases $d\geq2$, $R_n(\Delta_n) = o_p(n^{-1/2})$.
\end{corollary}

This is consistent with the geometric-quantile behavior reported by \cite{chaudhuri1996geometric, niemiro1992asymptotics}: the remainder vanishes for $d\geq2$, but the convergence rate is slower due to the logarithmic term in $d=2$. Consequently, under the conditions of Theorem \ref{prop:asynorm}, with $\mathbf A_0$ non-singular, the same first-order asymptotic distribution applies in the case $c=0$.



\section{Simulation study}\label{sec:sim}
In this section, we conduct a simulation study to evaluate the finite sample properties of the proposed method. This simulation exercise addresses the following issues. First, we consider different distributional choices for the error term to study its performance in the presence of non-Gaussian errors. Second, we evaluate the efficiency of the introduced estimator using different values of the tuning constant $c$. Here we focus on the bivariate case ($d=2$); additional simulation results with trivariate distributions and computational metrics are discussed in Section S1 of the Supplementary Materials.

To address the first aim of this simulation study, data are generated from the affine model:
\begin{equation*}
\bs Y = \bs A \mathbf{X} + \bs b,
\end{equation*} 
where the matrix $\bs A$ and the vector $\bs b$ are set as in \cite{he2017estimation}, i.e., $\bs A =\Bigl( \begin{smallmatrix} 2 & 0.3 \\ 0.3 & 1 \end{smallmatrix}\Bigr)$ and $\bs b = (3, 2)$. For the random vector $\bs X$, we consider the following scenarios:
\begin{enumerate}
\item[(N)]: a bivariate standard Normal distribution;
\item[(CN)]: a $\gamma$-contaminated bivariate Normal distribution $\bs X \sim (1-\gamma) \mathcal{N}_2 (\bs 0, \mathbb{I}) + \gamma \mathcal{N}_2 (\boldsymbol \mu, \bs \Sigma)$, where $\gamma$ is the proportion of outliers injected into the data. We fix $\boldsymbol \mu = (10, 10)$, $\bs \Sigma = 9 \bs A$ and a level of contamination $\gamma = 0.05$;
\item[(T)]: a bivariate Student-t distribution with 3 degrees of freedom.
\end{enumerate}

All scenarios consist of $n \in \{200, 500, 1000\}$ observations. To answer the second question, we compare the efficiency of the introduced estimator using three values for the tuning constant in \eqref{eq:rhohuber}, i.e., $c = 0$, $c = 1.5$ and an arbitrarily large value $c = 1000$. In the first and third case our model reduces, respectively, to a spatial-type and a geometric extremile formulation while the value $c=1.5$ in between down-weights the influence of outlying observations offering protection against outliers. The IRLS algorithm is run with a convergence tolerance set to $\epsilon = 10^{-8}$ and the maximum number of iterations to 1000.

Because of the symmetry property in Proposition \ref{prop:symmetry} we select $\tau \in \{0.05, 0.15, \dots, 0.45\}$ and consider 200 equispaced directions $\bs r_\theta = (\cos\theta, \sin\theta)$, $\theta \in [0,2\pi)$, covering the whole unit circle. For each scenario, we conduct $H=1000$ Monte Carlo replications and report the average and standard deviation values of the normalized squared error \citep{girard2017intriguing}:  
\begin{equation}
E_\tau = \tau \int_0^{2\pi}  \| \hat{\boldsymbol \xi}^c_{\tau, \bs r_\theta}(\bs Y)  -  \boldsymbol\xi^c_{\tau,\bs r_\theta} (\bs Y)  \|^2 \rd\theta,
\end{equation}
where the true extremile curves are obtained by solving the problem in \eqref{eq:rextrem} numerically. In the contaminated Normal scenario, the true extremile $\boldsymbol\xi^c_{\tau,\bs r_\theta} (\bs Y)$ is computed under the uncontaminated distribution. The \texttt{R} code for reproducing the results is available at \href{https://github.com/QuantileRegressionLab/MRExtremiles}{github.com/QuantileRegressionLab/MRExtremiles}.

Table \ref{tab:sim2d} reports the results for the considered sample sizes $n$, levels $\tau$, and tuning constants $c$. Consistently with the theoretical convergence results, in the Normal and Student-$t$ scenarios, the estimation error $E_\tau$ systematically decreases as the sample size increases from 200 to 1000. The numerical results further highlight the trade-off between robustness and efficiency controlled by the constant $c$. In the Normal scenario (Panel A), the geometric extremile estimator ($c=1000$) is the most efficient one, yielding the lowest average and standard deviation values. Conversely, in the presence of outliers (Panel B) or heavy tails (Panel C), it performs significantly worse with a non-vanishing large-sample error $E_\tau$, showing much higher error values and standard deviations compared to the spatial-like extremile ($c=0$) or robust extremile ($c=1.5$) versions. Specifically, in the contaminated case, the robust choice of $c$ effectively down-weights the influence of atypical observations, providing a stable estimation of the distribution's sparser regions.

\begin{table}
\centering
\begin{adjustbox}{width=0.7\textwidth}
\begin{tabular}{l l c c c c c}
\toprule
$n$ & $c$ & $\tau = 0.05$ & $\tau = 0.15$ & $\tau = 0.25$ & $\tau = 0.35$ & $\tau = 0.45$ \\
\midrule
\multicolumn{7}{c}{Panel A: Normal}\\
200  & 0    & $0.031 \; (0.014)$ & $0.035 \; (0.021)$ & $0.041 \; (0.030)$ & $0.047 \; (0.043)$ & $0.055 \; (0.059)$ \\
     & 1.5  & $0.025 \; (0.013)$ & $0.029 \; (0.019)$ & $0.034 \; (0.027)$ & $0.039 \; (0.039)$ & $0.046 \; (0.051)$ \\
     & 1000 & $0.024 \; (0.013)$ & $0.027 \; (0.019)$ & $0.031 \; (0.026)$ & $0.035 \; (0.035)$ & $0.041 \; (0.044)$ \\
500  & 0    & $0.013 \; (0.006)$ & $0.014 \; (0.008)$ & $0.017 \; (0.012)$ & $0.019 \; (0.017)$ & $0.022 \; (0.023)$ \\
     & 1.5  & $0.010 \; (0.005)$ & $0.012 \; (0.008)$ & $0.014 \; (0.011)$ & $0.016 \; (0.015)$ & $0.019 \; (0.020)$ \\
     & 1000 & $0.010 \; (0.005)$ & $0.011 \; (0.008)$ & $0.013 \; (0.010)$ & $0.015 \; (0.014)$ & $0.017 \; (0.018)$ \\
1000 & 0    & $0.007 \; (0.003)$ & $0.007 \; (0.004)$ & $0.009 \; (0.006)$ & $0.010 \; (0.009)$ & $0.011 \; (0.012)$ \\
     & 1.5  & $0.005 \; (0.003)$ & $0.006 \; (0.004)$ & $0.007 \; (0.006)$ & $0.008 \; (0.008)$ & $0.010 \; (0.010)$ \\
     & 1000 & $0.005 \; (0.003)$ & $0.006 \; (0.004)$ & $0.007 \; (0.005)$ & $0.008 \; (0.007)$ & $0.009 \; (0.009)$ \\
           \\
\multicolumn{7}{c}{Panel B: Contaminated Normal}\\
200  & 0    & $1.279 \; (1.760)$ & $0.102 \; (0.067)$ & $0.082 \; (0.060)$ & $0.079 \; (0.071)$ & $0.086 \; (0.092)$ \\
     & 1.5  & $1.299 \; (1.719)$ & $0.116 \; (0.079)$ & $0.084 \; (0.064)$ & $0.078 \; (0.071)$ & $0.084 \; (0.087)$ \\
     & 1000 & $3.123 \; (1.528)$ & $1.635 \; (0.908)$ & $1.076 \; (0.619)$ & $0.810 \; (0.486)$ & $0.780 \; (0.504)$ \\
500  & 0    & $0.843 \; (1.095)$ & $0.073 \; (0.041)$ & $0.053 \; (0.034)$ & $0.048 \; (0.039)$ & $0.051 \; (0.049)$ \\
     & 1.5  & $0.882 \; (1.063)$ & $0.091 \; (0.049)$ & $0.059 \; (0.038)$ & $0.051 \; (0.040)$ & $0.054 \; (0.048)$ \\
     & 1000 & $3.035 \; (1.058)$ & $1.546 \; (0.620)$ & $1.006 \; (0.420)$ & $0.750 \; (0.329)$ & $0.735 \; (0.344)$ \\
1000 & 0    & $0.607 \; (0.586)$ & $0.062 \; (0.027)$ & $0.042 \; (0.022)$ & $0.036 \; (0.024)$ & $0.038 \; (0.030)$ \\
     & 1.5  & $0.659 \; (0.566)$ & $0.080 \; (0.032)$ & $0.050 \; (0.025)$ & $0.041 \; (0.025)$ & $0.042 \; (0.030)$ \\
     & 1000 & $2.982 \; (0.717)$ & $1.497 \; (0.413)$ & $0.968 \; (0.278)$ & $0.718 \; (0.216)$ & $0.707 \; (0.227)$ \\
           \\
\multicolumn{7}{c}{Panel C: Student-t}\\
200  & 0    & $0.109 \; (0.075)$ & $0.065 \; (0.043)$ & $0.058 \; (0.043)$ & $0.058 \; (0.052)$ & $0.064 \; (0.067)$ \\
     & 1.5  & $0.100 \; (0.073)$ & $0.060 \; (0.043)$ & $0.054 \; (0.043)$ & $0.055 \; (0.051)$ & $0.061 \; (0.065)$ \\
     & 1000 & $0.356 \; (1.225)$ & $0.182 \; (0.399)$ & $0.134 \; (0.227)$ & $0.114 \; (0.164)$ & $0.116 \; (0.151)$ \\
500  & 0    & $0.043 \; (0.024)$ & $0.026 \; (0.016)$ & $0.023 \; (0.017)$ & $0.023 \; (0.020)$ & $0.025 \; (0.026)$ \\
     & 1.5  & $0.040 \; (0.023)$ & $0.024 \; (0.016)$ & $0.021 \; (0.017)$ & $0.022 \; (0.021)$ & $0.024 \; (0.026)$ \\
     & 1000 & $0.150 \; (0.190)$ & $0.076 \; (0.074)$ & $0.055 \; (0.050)$ & $0.047 \; (0.044)$ & $0.048 \; (0.050)$ \\
1000 & 0    & $0.021 \; (0.011)$ & $0.012 \; (0.007)$ & $0.011 \; (0.008)$ & $0.011 \; (0.010)$ & $0.012 \; (0.013)$ \\
     & 1.5  & $0.019 \; (0.011)$ & $0.011 \; (0.007)$ & $0.010 \; (0.008)$ & $0.011 \; (0.010)$ & $0.012 \; (0.013)$ \\
     & 1000 & $0.071 \; (0.102)$ & $0.035 \; (0.035)$ & $0.026 \; (0.023)$ & $0.022 \; (0.020)$ & $0.022 \; (0.022)$ \\
\bottomrule
\end{tabular}
\end{adjustbox}
\caption{Average and standard deviation values of the normalized squared error $E_\tau$ for all considered cases.}
\label{tab:sim2d}
\end{table}

\begin{figure}[!h]
    \centering
    \includegraphics[width=1\textwidth]{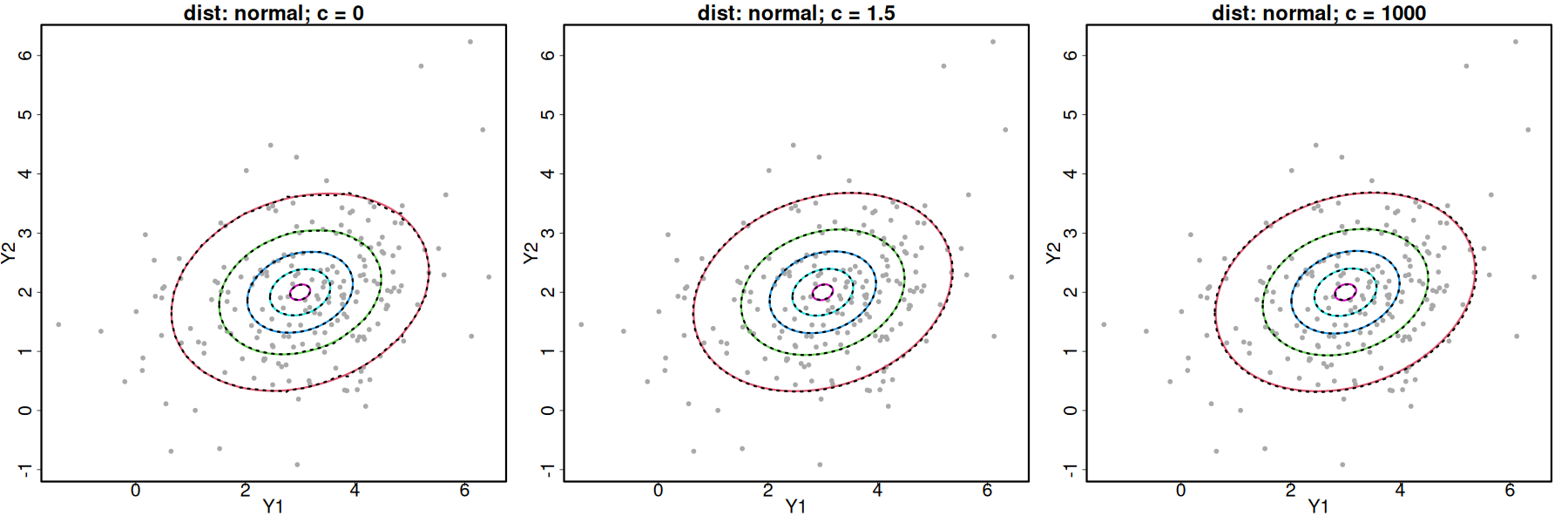}
    \includegraphics[width=1\textwidth]{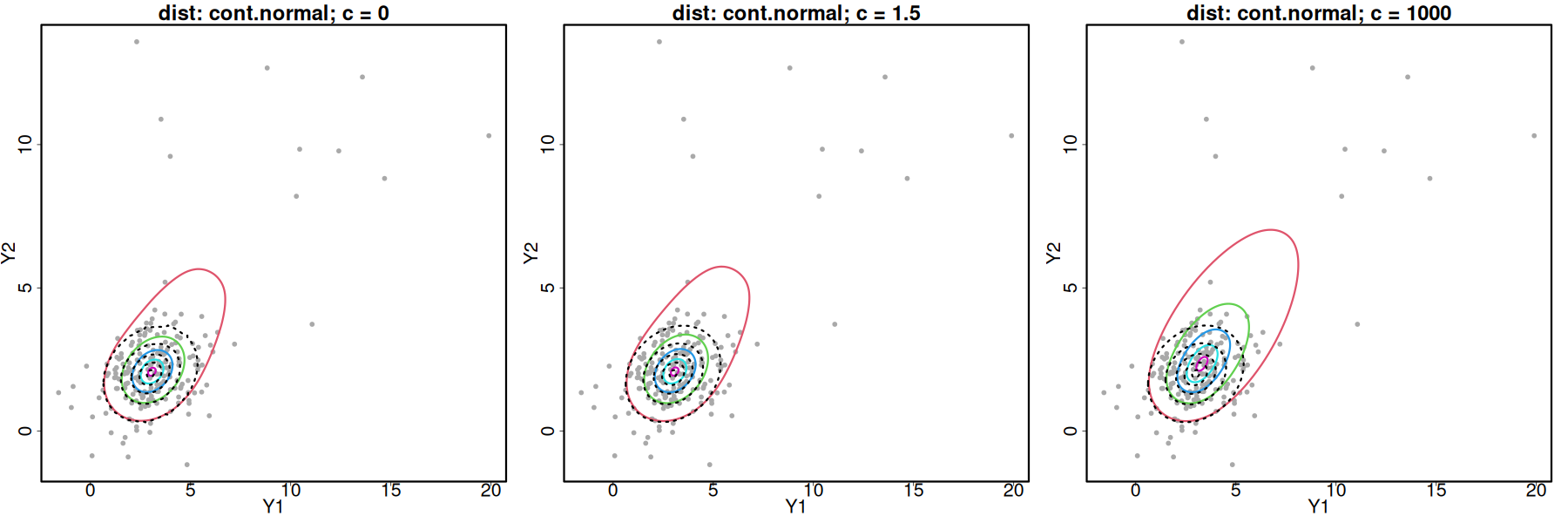}
    \includegraphics[width=1\textwidth]{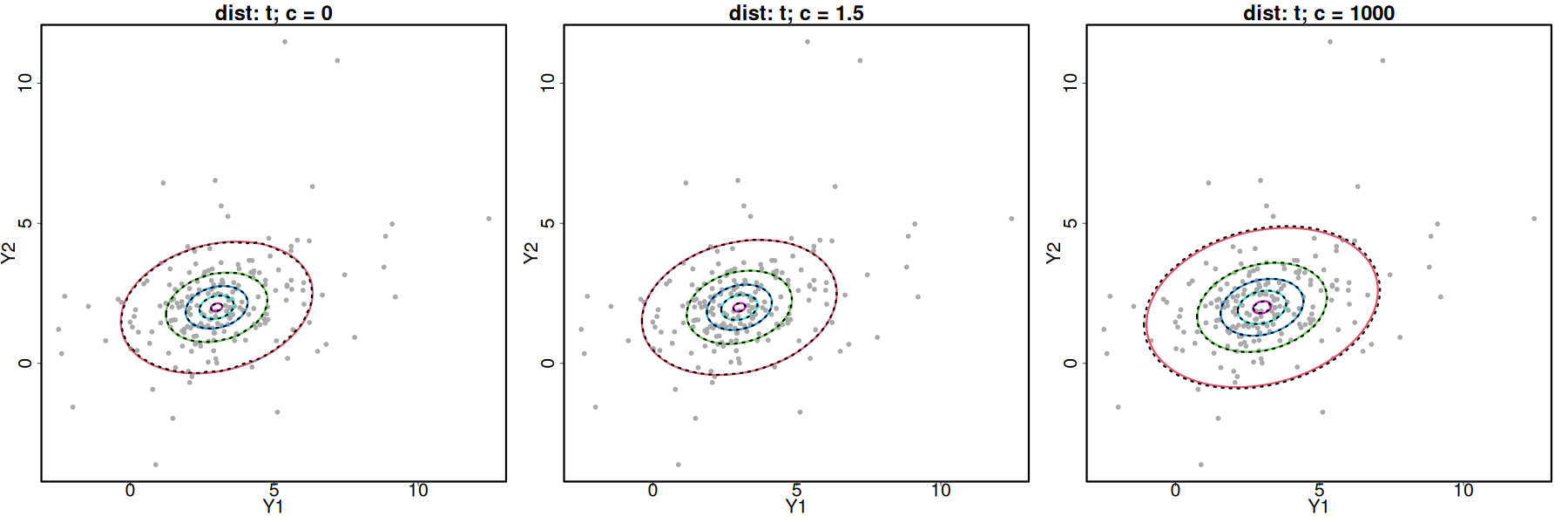}
    \caption{True (dashed black lines) and estimated (solid lines) extremile contours with $c=0$ (first column), $c=1.5$ (second column) and $c=1000$ (third column) under the normal (first row), contaminated normal (second row) and the Student-t (third row) simulation scenarios at levels $\tau$, 0.05 (red), 0.15 (green), 0.25 (dark blue), 0.35 (light blue) and 0.45 (violet). Data are illustrated as gray points.}
    \label{fig:sim_cont}
\end{figure}

To get an impression of how the proposed multivariate extremiles behave in the above scenarios, Figure \ref{fig:sim_cont} overlays the true (dashed lines) and estimated (solid lines) directional M-extremile curves averaged over the $H$ replicates for the considered values of $\tau$ and $c$ with sample size $n = 200$. The visual evidence confirms that the estimated regions closely track the true ones across all considered levels of $\tau$, even with a relatively small sample size. As $\tau$ decreases, in the considered settings the contours are nested and move toward the outer regions of the support, characterizing more sparse spaces. One also observes that the regions tend to contract with increasing $c$ due to the fact that a larger value for $c$ means that outliers have a stronger influence. The second row of Figure \ref{fig:sim_cont} visually validates the robustness findings from Table \ref{tab:sim2d}: while the contours for $c=1000$ are visibly distorted by the injected outliers, the contours for $c=1.5$ and $c=0$ remain robustly centered around the main data mass.

\section{Application}\label{sec:app}
We apply the proposed method to daily returns of three major cryptocurrencies by market capitalization: Bitcoin (BTC), Ethereum (ETH), and Binance Coin (BNB). The use of extremile regions is particularly natural in risk management, where characterizing the magnitude of extreme joint losses is critical for informed investment decisions. To ensure the reproducibility of our findings, the full analysis and associated code are available in our GitHub repository.

The dataset consists of daily observations retrieved from Yahoo Finance (\href{https://finance.yahoo.com}{finance.yahoo.com}) spanning from March 2021 to March 2026. Daily returns with continuous compounding are calculated by taking the logarithm of the ratio of closing prices on consecutive trading days. The final sample size consists of $n=1826$ returns.

It is well known that financial data are characterized by stylized facts like heavy tails, skewness, and the presence of outliers. Preliminary analysis confirms these features in our data, with Ljung-Box tests rejecting the null hypothesis of serial independence for all three assets. Since the raw returns violate the i.i.d. assumption, we follow the solution of \cite{he2017estimation} and perform our analysis on standardized residuals, also called \qmo innovations\qmc, extracted by fitting a univariate exponential GARCH(1,1) \citep{nelson1991conditional} to each series assuming generalized error distribution errors. We then tested whether the standardized absolute and squared residuals were serially correlated using Ljung-Box tests at the 1\% significance level. The tests provided no evidence of remaining serial dependence in the absolute or squared standardized residuals at the 1\% significance level. 


In the following, our objective is to construct multivariate extremile sets for the joint distribution of the standardized innovations $\bs Z = (Z_1^{\rm BTC}, Z_2^{\rm ETH}, Z_3^{\rm BNB})$. To capture unexpected high losses and account for outliers, we implement the proposed approach at level $\tau = 0.005$ with $c=0$, evaluating the regions across uniformly distributed directions over the unit sphere $\mathcal{S}^2$.

\begin{sidewaysfigure}[!h]
    \centering
    \includegraphics[width=0.33\textwidth]{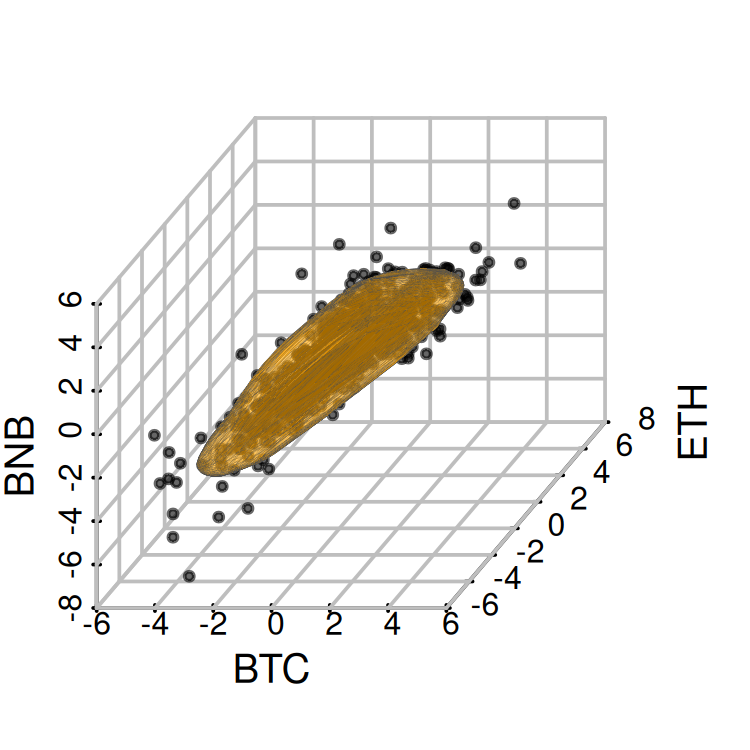}
    \includegraphics[width=0.33\textwidth]{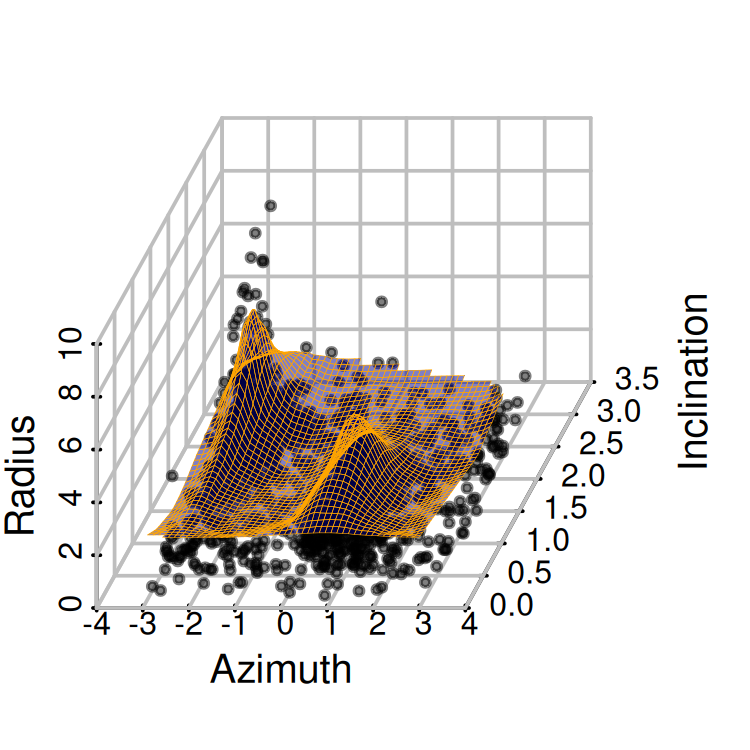}
    \includegraphics[width=0.33\textwidth]{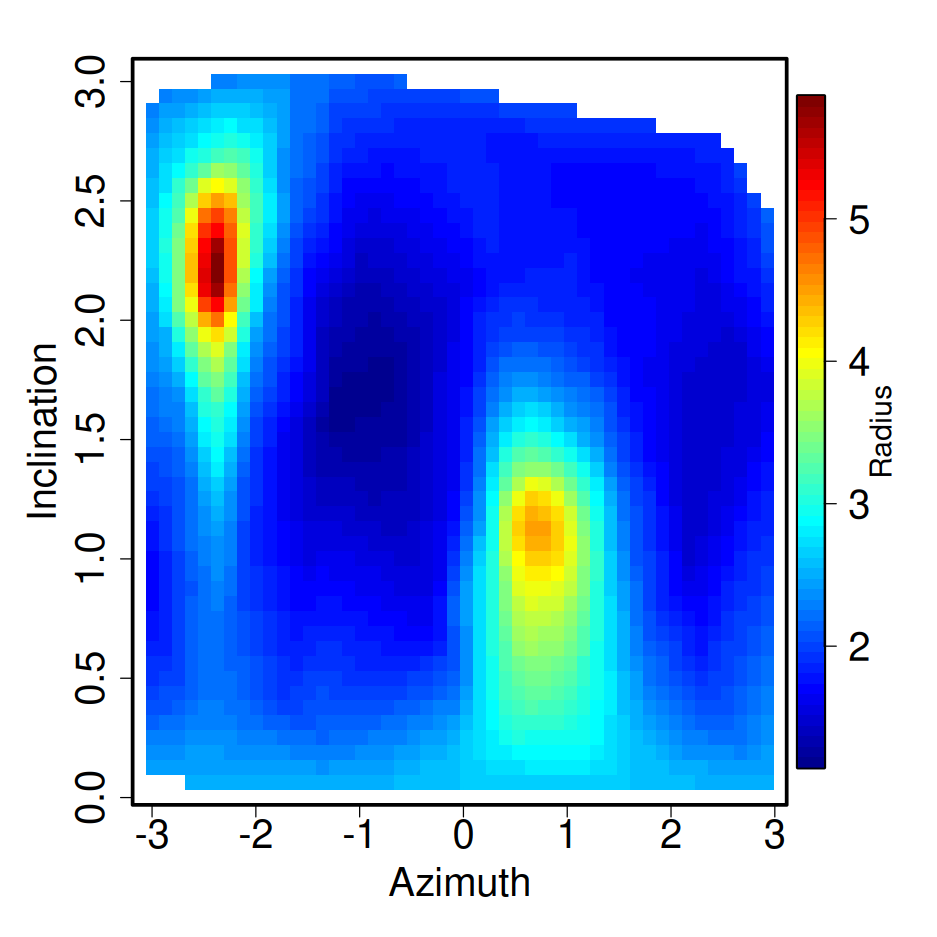}
    \caption{Left: estimated trivariate extremile region; Right: spherical-coordinate representation; Bottom: heatmap distributions of spherical coordinates.}
    \label{fig:res3D}
\end{sidewaysfigure}

Figure \ref{fig:res3D} summarizes the main empirical findings through three complementary plots: the estimated trivariate extremile region (left), its spherical-coordinate representation (right), and the corresponding heatmap of the spherical-coordinate distribution (bottom). The left 3D plot shows that the estimated extremile region is markedly elongated rather than spherical. This indicates that the overall shape is driven primarily by joint movements of the three assets. 

The spherical-coordinate representations in Figure \ref{fig:res3D} provide a second layer of interpretation. In the right plot, the orange surface describes the radial size of the extremile contour as a function of the direction. Two pronounced peaks emerge, separated by flatter regions in between where the contour remains much closer to the center. That means the extremile contour is highly anisotropic. Taken together with the 3D contour, these dominant directions correspond to simultaneous shocks across the three cryptocurrencies, which would indicate strong joint downside risk.

Finally, the heatmap reinforces these conclusions by revealing two localized hot regions, corresponding to the directions in which the extremile radius is largest. These peaks are not symmetric, either in intensity or in shape. In terms of intensity, the darkest red region indicates that combinations of negative BTC–ETH–BNB shocks are considerably more extreme than those associated with positive joint returns, represented by the weaker yellow-orange region. In terms of shape, the less intense peak appears more irregular and suggests that, although the dominant extreme pattern is still associated with common movements, this secondary mode is also influenced by more asset-specific variation (particularly for BNB). 

\section{Conclusions}\label{sec:con}
This paper introduces a new class of robust multivariate extremiles by extending the univariate extremile of \cite{daouia2019extremiles} to the multivariate setting using a directional approach.

The proposed method preserves the main feature of extremiles while explicitly incorporating robustness through Huber's multidimensional loss function, thereby yielding a flexible family of multivariate tail functionals that bridges quantile-like and expectile-like behavior. In this sense, the first main novelty of the paper is methodological: to the best of our knowledge, this is the first (robust) extension of extremiles and extremile sets associated with random vectors. A second major contribution is theoretical. The paper establishes existence and uniqueness of the proposed functional under mild assumptions, together with key desirable properties. In addition, the proposed methodology is computationally efficient and easy to implement using IRLS in a two-step procedure. 
From an inferential point of view, the empirical estimator is framed as a Z-estimator, which allowed us to derive consistency and asymptotic normality. The simulation study highlights the practical value of the robustness parameter $c$ under Gaussian data, and in the presence of contamination or heavy tails. The empirical application to standardized innovations of Bitcoin, Ethereum, and Binance Coin illustrates the practical relevance of the approach in a financial context. The shape and orientation of the estimated trivariate extremile region show that joint tail risk in crypto markets is highly anisotropic and primarily driven by a limited set of common shocks rather than by isolated idiosyncratic movements.

Several directions for future research remain open. 
First, the proposed robust extremile requires choosing a reasonable value for the tuning constant, based on the data structure and the aim of the analysis. For instance, a practical strategy to choose $c$ is to attain a target asymptotic relative efficiency, such as 95\%, relative to the sample mean under a multivariate Gaussian reference distribution, averaged over directions at $\tau = 1/2$. Second, the approach we propose is clearly not the only possible extension of extremiles to a multivariate setting. Alternative specifications such as those based on joint distribution functions or copula formulations can be considered. On the applied side, extensions to conditional multivariate extremiles and regression settings \citep{daouia2022extremile} appear particularly promising.

\bigskip
\begin{center}
{\bf SUPPLEMENTARY MATERIALS}
\end{center}

\begin{description}
\item[Simulations, computational metrics and proofs:] Simulation studies, computational metrics and lengthy technical derivations that are used to support the results in the manuscript. (PDF file)
\end{description}

\begin{center}
{\bf DATA AVAILABILITY STATEMENT}
\end{center}

\begin{description}
\item[Data and code:] The data that support the findings of this study are openly available in Yahoo Finance and the code can be found at \href{https://github.com/QuantileRegressionLab/MRExtremiles}{github.com/QuantileRegressionLab/MRExtremiles}.
\end{description}

\bibliographystyle{agsm}
\bibliography{bibliography}

@article{daouia2019extremiles,
  title={Extremiles: A new perspective on asymmetric least squares},
  author={Daouia, Abdelaati and Gijbels, Ir{\`e}ne and Stupfler, Gilles},
  journal={Journal of the American Statistical Association},
  volume={114},
  number={527},
  pages={1366--1381},
  year={2019},
  publisher={Taylor \& Francis}
}

@article{chaudhuri1996geometric,
  title={On a geometric notion of quantiles for multivariate data},
  author={Chaudhuri, Probal},
  journal={Journal of the American Statistical Association},
  volume={91},
  number={434},
  pages={862--872},
  year={1996},
  publisher={Taylor \& Francis}
}

@article{maume2017multivariate,
  author  = {Maume-Deschamps, V{\'e}ronique and Rulliere, Didier and Said, Khalil},
  title   = {Multivariate extensions of expectiles risk measures},
  journal = {Dependence Modeling},
  volume  = {5},
  number  = {1},
  pages   = {20--44},
  year    = {2017}
}

@article{he1997convergence,
  title={Convergence of depth contours for multivariate datasets},
  author={He, Xuming and Wang, Gang},
  journal={The Annals of Statistics},
  pages={495--504},
  year={1997},
  volume={25},
  number={2},
  publisher={JSTOR}
}

@article{kong2012quantile,
  title={Quantile tomography: using quantiles with multivariate data},
  author={Kong, Linglong and Mizera, Ivan},
  journal={Statistica Sinica},
  pages={1589--1610},
  year={2012},
  volume={22},
  number={4},
  publisher={JSTOR}
}

@article{hallin2010multivariate,
  title={MULTIVARIATE QUANTILES AND MULTIPLE-OUTPUT REGRESSION QUANTILES: FROM ${L}_{1}$ OPTIMIZATION TO HALFSPACE DEPTH},
  author={Hallin, Marc and Paindaveine, Davy and {\v{S}}iman, Miroslav and Wei, Ying and Serfling, Robert and Zuo, Yijun and Kong, Linglong and Mizera, Ivan},
  journal={The Annals of Statistics},
  pages={635--703},
  year={2010},
  publisher={JSTOR}
}

@article{breckling1988m,
  title={M-quantiles},
  author={Breckling, Jens and Chambers, Ray},
  journal={Biometrika},
  pages={761--771},
  year={1988},
  publisher={JSTOR}
}

@incollection{kokic2002new,
  title={A new definition of multivariate {M}-quantiles},
  author={Kokic, Philip and Breckling, Jens and L{\"u}bke, Oliver},
  booktitle={Statistical data analysis based on the L$_1$-norm and related methods},
  pages={15--24},
  year={2002},
  publisher={Springer}
}

@book{van2000asymptotic,
  title={Asymptotic {S}tatistics},
  author={Van der Vaart, Aad W},
  volume={3},
  year={2000},
  publisher={Cambridge University Press}
}

@article{newey1987asymmetric,
  title={Asymmetric least squares estimation and testing},
  author={Newey, Whitney K and Powell, James L},
  journal={Econometrica: Journal of the Econometric Society},
  pages={819--847},
  year={1987},
  volume={55},
  number={4},
  publisher={JSTOR}
}

@article{bellini2014generalized,
  title={Generalized quantiles as risk measures},
  author={Bellini, Fabio and Klar, Bernhard and M{\"u}ller, Alfred and Gianin, Emanuela Rosazza},
  journal={Insurance: Mathematics and Economics},
  volume={54},
  pages={41--48},
  year={2014},
  publisher={Elsevier}
}

@article{delbaen2016risk,
  title={Risk measures with the {CxLS} property},
  author={Delbaen, Freddy and Bellini, Fabio and Bignozzi, Valeria and Ziegel, Johanna F},
  journal={Finance and Stochastics},
  volume={20},
  number={2},
  pages={433--453},
  year={2016},
  publisher={Springer}
}

@article{acerbi2002spectral,
  title={Spectral measures of risk: A coherent representation of subjective risk aversion},
  author={Acerbi, Carlo},
  journal={Journal of Banking \& Finance},
  volume={26},
  number={7},
  pages={1505--1518},
  year={2002},
  publisher={Elsevier}
}

@article{wang1996premium,
  title={Premium calculation by transforming the layer premium density},
  author={Wang, Shaun},
  journal={ASTIN Bulletin: The Journal of the IAA},
  volume={26},
  number={1},
  pages={71--92},
  year={1996},
  publisher={Cambridge University Press}
}

@article{geng2024modelling,
  title={Modelling additive extremile regression by iteratively penalized least asymmetric weighted squares and gradient descent boosting},
  author={Geng, Ziwen},
  journal={Statistics},
  volume={58},
  number={3},
  pages={576--595},
  year={2024},
  publisher={Taylor \& Francis}
}

@book{hampel2011robust,
  title={Robust statistics: the approach based on influence functions},
  author={Hampel, Frank R and Ronchetti, Elvezio M and Rousseeuw, Peter J and Stahel, Werner A},
  volume={196},
  year={2011},
  publisher={John Wiley \& Sons}
}

@article{nelson1991conditional,
  title={Conditional heteroskedasticity in asset returns: A new approach},
  author={Nelson, Daniel B},
  journal={Econometrica: Journal of the Econometric Society},
  pages={347--370},
  year={1991},
  volume={59},
  number={2},
  publisher={JSTOR}
}

@article{debrauwer2024generalized,
  title={Generalized extremiles and risk measures of distorted random variables},
  author={Debrauwer, Dieter and Gijbels, Ir{\`e}ne and Herrmann, Klaus},
  journal={arXiv:2405.11248},
  year={2024}
}

@article{sun2026robust,
  title={Robust extremile regression in high dimensions with {H}uber loss},
  author={Sun, Weixi and Wang, Shanshan},
  journal={Statistical Papers},
  volume={67},
  number={4},
  pages={89},
  year={2026},
  publisher={Springer}
}

@article{konen2022multivariate,
  title={Multivariate $\rho$-quantiles: A spatial approach},
  author={Konen, Dimitri and Paindaveine, Davy},
  journal={Bernoulli},
  volume={28},
  number={3},
  pages={1912--1934},
  year={2022},
  publisher={Bernoulli Society for Mathematical Statistics and Probability}
}

@article{serfling2002quantile,
  title={Quantile functions for multivariate analysis: approaches and applications},
  author={Serfling, Robert},
  journal={Statistica Neerlandica},
  volume={56},
  number={2},
  pages={214--232},
  year={2002},
  publisher={Wiley Online Library}
}

@article{daouia2022extremile,
  title={Extremile regression},
  author={Daouia, Abdelaati and Gijbels, Irene and Stupfler, Gilles},
  journal={Journal of the American Statistical Association},
  volume={117},
  number={539},
  pages={1579--1586},
  year={2022},
  publisher={Taylor \& Francis}
}

@article{fissler2016higher,
  title={Higher order elicitability and {O}sband's principle},
  author={Fissler, Tobias and Ziegel, Johanna F},
  journal={The Annals of Statistics},
  pages={1680--1707},
  year={2016},
  publisher={JSTOR}
}

@article{artzner1999coherent,
  title={Coherent measures of risk},
  author={Artzner, Philippe and Delbaen, Freddy and Eber, Jean-Marc and Heath, David},
  journal={Mathematical Finance},
  volume={9},
  number={3},
  pages={203--228},
  year={1999},
  publisher={Wiley Online Library}
}

@article{merlo2024unified,
  title={Unified unconditional regression for multivariate quantiles, M-quantiles, and expectiles},
  author={Merlo, Luca and Petrella, Lea and Salvati, Nicola and Tzavidis, Nikos},
  journal={Journal of the American Statistical Association},
  volume={119},
  number={547},
  pages={2154--2165},
  year={2024},
  publisher={Taylor \& Francis}
}

@article{he2017estimation,
  title={Estimation of extreme depth-based quantile regions},
  author={He, Yi and Einmahl, John HJ},
  journal={Journal of the Royal Statistical Society Series B: Statistical Methodology},
  volume={79},
  number={2},
  pages={449--461},
  year={2017},
  publisher={Oxford University Press}
}

@article{girard2017intriguing,
  title={Intriguing properties of extreme geometric quantiles},
  author={Girard, St{\'e}phane and Stupfler, Gilles},
  journal={REVSTAT-Statistical Journal},
  volume={15},
  number={1},
  pages={107--139},
  year={2017}
}

@article{herrmann2018multivariate,
  title={Multivariate geometric expectiles},
  author={Herrmann, Klaus and Hofert, Marius and Mailhot, M{\'e}lina},
  journal={Scandinavian Actuarial Journal},
  volume={2018},
  number={7},
  pages={629--659},
  year={2018},
  publisher={Taylor \& Francis}
}

@article{bahadur1966note,
  title={A note on quantiles in large samples},
  author={Bahadur, R Raj},
  journal={The Annals of Mathematical Statistics},
  volume={37},
  number={3},
  pages={577--580},
  year={1966},
  publisher={JSTOR}
}

@article{niemiro1992asymptotics,
  title={Asymptotics for {M}-estimators defined by convex minimization},
  author={Niemiro, Wojciech and others},
  journal={The Annals of Statistics},
  volume={20},
  number={3},
  pages={1514--1533},
  year={1992},
  publisher={Institute of Mathematical Statistics}
}

@article{dyckerhoff2011weighted,
  title={Weighted-mean trimming of multivariate data},
  author={Dyckerhoff, Rainer and Mosler, Karl},
  journal={Journal of Multivariate Analysis},
  volume={102},
  number={3},
  pages={405--421},
  year={2011},
  publisher={Elsevier}
}


\end{document}